\documentclass[a4paper]{jpconf}
\usepackage{iopams}
\usepackage{amsmath}
\usepackage{amsfonts}
\usepackage{multirow} 
\usepackage{graphicx} 

\begin{document}

\vspace{-0.2cm}
\hbox{DO-TH 11/02}

\title{Angular analysis of $B \to V (\to P_1 P_2) + \bar\ell\ell$ decays}

\author{Christoph Bobeth$^{1,}$\footnote[3]{Speaker}, Gudrun Hiller$^2$ and Danny
  van Dyk$^2$\\[2mm]}

\address{$^1$ Institute of Advanced Study \& Excellence Cluster Universe, 
  Technische Universit\"at M\"unchen, D-85748 Garching, Germany \\[2mm]}

\address{$^2$ Institut f\"ur Physik, Technische Universit\"at Dortmund, D-44221
  Dortmund, Germany \\[2mm]}

\ead{christoph.bobeth@ph.tum.de}

\begin{abstract}
  The angular analysis of exclusive rare $B$-meson decays via intermediate
  vector mesons $V$ into 4-body final states of two pseudo-scalars $P_1, P_2$
  and a pair of light leptons $\ell = e, \mu$ offers a large set of
  observables. They can be used to test the electroweak short-distance couplings
  in the Standard Model and to search for New Physics. The two kinematic regions
  of low and high dilepton mass depend on short-distance physics in
  complementary ways and can be expanded in powers of $\Lambda_{\rm
    QCD}/m_b$. These expansions guide towards suitable combinations of
  observables allowing to $i)$ reduce the hadronic uncertainties in the
  extraction of the short-distance couplings or $ii)$ test the lattice QCD $B
  \to V$ form factors in short-distance independent combinations. Several such
  possibilities of CP-averaged and CP-asymmetric (T-even and T-odd) quantities
  are presented for $\bar{B}_d^0 \to \bar{K}^{*0} (\to K^- \pi^+) +
  \bar\ell\ell$ and time-integrated CP-asymmetries without tagging for
  $\bar{B}_s,B_s \to \phi (\to K^- K^+) + \bar\ell\ell$ decays in view of the
  latest $B$-factory and CDF results and the forthcoming LHCb measurements.
\end{abstract}

%
%
\section{Introduction}

\vskip0.3cm

The $b \to s + \bar\ell\ell$ induced flavour changing neutral current (FCNC)
decays of the $B$-meson are well known to provide sensitive probes of the
electroweak short-distance couplings of the Standard Model (SM) and to test
scenarios beyond (BSM). Hitherto existing experimental results from the
$B$-factory experiments Babar~\cite{Aubert:2008ju} and Belle~\cite{Wei:2009zv}
as well as the Tevatron experiment CDF~\cite{Aaltonen:2011cn} are in the
ballpark of the SM predictions and provide constraints on the couplings ${\cal
  C}_{9,10}$ related to 4-Fermi operators $(\bar{s} \gamma_\mu P_L
b)\,(\bar{\ell} \gamma^\mu \{1, \gamma_5\} \ell)$ \cite{Bobeth:2010wg,
  Bobeth:2011gi}. Further experimental progress is expected soon from the final
analysis of the full data sets of Babar and Belle, as well as the CDF one with
at least doubled statistics. Eventually, LHCb will dominate statistically for
final states containing charged particles only, already with a rather low
luminosity of a few fb$^{-1}$~\cite{:2009ny}. The Super-$B$ factories are
expected to contribute to $\ell = e$, just as the $B$-factories do presently.

\begin{figure}
\begin{minipage}[b]{0.50\textwidth}
  \includegraphics[width=8.0cm, angle=0]{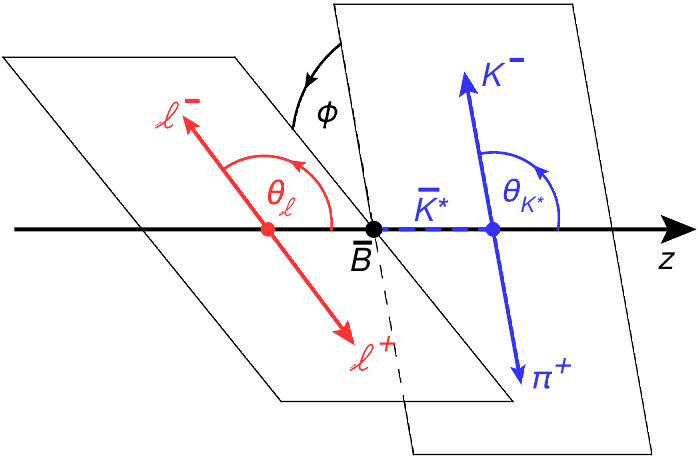} 
\end{minipage}
\hspace{0.03\textwidth}
\begin{minipage}[b]{0.40\textwidth}
  \caption{\label{fig:4bdy:kin} Kinematic variables of $\bar{B}_d^0 \to
    \bar{K}^{*0} (\to K^- \pi^+) + \bar\ell\ell$ decays: \\
    $i)$ the $(\bar\ell\ell)$-invariant mass squared~$q^2$,
    $ii)$ the angle $\theta_\ell$ between $\ell = \ell^-$ and $\bar{B}$ in the
    $(\bar\ell\ell)$ center of mass (c.m.), 
    $iii)$ the angle $\theta_{K^*}$ between $K^-$ and $\bar{B}$ in the
    $(K^-\pi^+)$ c.m. and
    $iv)$ the angle $\phi$ between the two decay planes spanned by the 3-momenta
    of the $(K\pi)$- and $(\bar\ell\ell)$-systems, respectively. }
\end{minipage}
\end{figure}

Especially the angular analysis of the 4-body final state of $B \to V (\to P_1
P_2) + \bar\ell\ell$ decays offers a large number of observables in the fully
differential distribution \cite{Kruger:1999xa, Kim:2000dq}. Here the
intermediate $V$ is assumed to be on-shell in the narrow-resonance approximation
which restricts the number of kinematic variables to four\footnote{The
  off-resonance case has been studied in \cite{Grinstein:2005ud}.}. Using
$\bar{B}_d^0 \to \bar{K}^{*0} (\to K^- \pi^+) + \bar\ell\ell$ for illustration,
they might be chosen as depicted in figure \ref{fig:4bdy:kin}.

The differential decay rate, after summing over lepton spins, factorises into
\begin{align}
  \frac{8 \pi}{3} \frac{d^4\Gamma}{dq^2\, d\cos\theta_\ell\, d\cos\theta_{K^*}\, d\phi} 
  = J_1^s \sin^2\theta_{K^*} + J_1^c \cos^2\theta_{K^*}
  + (J_2^s \sin^2\theta_{K^*} + J_2^c \cos^2\theta_{K^*}) \cos 2\theta_\ell
\nonumber \\
  + J_3 \sin^2\theta_{K^*} \sin^2\theta_\ell \cos 2\phi  
  + J_4 \sin 2\theta_{K^*} \sin 2\theta_\ell \cos\phi 
  + J_5 \sin 2\theta_{K^*} \sin\theta_\ell \cos\phi
\nonumber \\[0.2cm]  
  + (J_6^s \sin^2\theta_{K^*} + J_6^c \cos^2\theta_{K^*}) \cos\theta_\ell 
  + J_7 \sin 2\theta_{K^*} \sin\theta_\ell \sin\phi
\nonumber \\[0.2cm] 
  + J_8 \sin 2\theta_{K^*} \sin 2\theta_\ell \sin\phi
  + J_9 \sin^2\theta_{K^*} \sin^2\theta_\ell \sin 2\phi, 
\end{align}
that is, into $q^2$-dependent observables\footnote{Possibilities to extract
  $q^2$-integrated $J_i^{j}$ from single-differential distributions in
  $\theta_\ell$, $\theta_{K^*}$ or $\phi$ can be found in~\cite{Bobeth:2008ij}.}
$J_i^{j}(q^2)$ and the dependence on the angles $\theta_\ell$, $\theta_{K^*}$
and $\phi$. No additional angular dependencies can be induced by any extension
of the SM operator basis~\cite{Bobeth:1999mk} as found by
\cite{Altmannshofer:2008dz, Alok:2010zd}.  The following simplifications arise
in the limit $m_\ell \to 0$: $J_1^s = 3 J_2^s$, $J_1^c = - J_2^c$ and $J_6^c =
0$.

The differential decay rate $d^4\bar{\Gamma}$ of the CP-conjugated decay $B_d^0
\to K^{0*} (\to K^+ \pi^-) + \bar\ell\ell$ is obtained through the following
replacements
\begin{align}
  J_{1,2,3,4,7}^{j} & \to \, \bar{J}_{1,2,3,4,7}^{j}[\delta_W \to -\delta_W], &
  J_{5,6,8,9}^{j} & \to -\, \bar{J}_{5,6,8,9}^{j}[\delta_W \to -\delta_W], &
\end{align}
due to $\ell \leftrightarrow \bar\ell \Rightarrow \theta_\ell \to \theta_\ell -
\pi$ and $\phi \to -\phi$. The CP-violating (weak) phases $\delta_W$ are
conjugated.

The angular distribution provides twice as many observables ($J_i^j$ and
$\bar{J}_i^j$) when the decay and its CP-conjugate decay are measured
separately. This doubles again if the $\ell = e$ and $\mu$ lepton flavours are
not averaged.  Notably, CP-asymmetries can be measured in an untagged sample of
$B$-mesons due to the presence of CP-odd observables ($i = 5,6,8,9$)
\cite{Kruger:1999xa}. Moreover, T-odd observables $\sim \cos \delta_s
\sin\delta_W$ ($i = 7,8,9$) are especially sensitive to weak BSM phases
$\delta_W$ \cite{Bobeth:2008ij, Chen:2002bq} contrary to T-even ones $\sim \sin
\delta_s \sin\delta_W$ ($i=1,\ldots,6$), since the CP-conserved (strong) phase
$\delta_s$ is often predicted to be small. Note, that in the SM CP-violating
effects in $b\to s$ are doubly-suppressed by the Cabibbo angle as ${\rm Im}
[V_{ub}^{} V_{us}^*/(V_{tb}^{} V_{ts}^*)] \approx \bar{\eta} \lambda \sim
10^{-2}$.

The observables $J_i^j$ are bilinear in the transversity amplitudes $A_a^{L,R}$
($a = 0, \perp, \parallel$ in the limit $m_\ell \to 0$\footnote{Two more
  amplitudes contribute for $m_\ell \neq 0$: $A_t$ (time-like) and in the
  presence of scalar operators, $A_S$ \cite{Altmannshofer:2008dz}.})
\cite{Kruger:2005ep}. Here $L,R$ refer to the chirality of the lepton current.
The kinematic dependence on the lepton mass is suppressed by
$m_\ell^2/q^2$. Hence, non-negligible effects from finite lepton masses at $q^2 >
1$~GeV$^2$ arise only in BSM scenarios, as for example in the MSSM at large
$\tan\beta$ due to neutral Higgs boson exchange, see for instance
\cite{Bobeth:2007dw}.

Depending on $q^2$, present theoretical predictions suffer from long-distance
dominated $(\bar{q}q)$-resonance background ($q = u,d,s,c$) induced by
current-current- and QCD-penguin-operators of the $\Delta B = 1$ effective
Hamiltonian. Compared to the leading $c$-current-current contribution, the
$u$-current-current one is suppressed by small CKM elements and the QCD-penguin
contribution by small Wilson coefficients. The light ($u,d,s$)-resonances affect
the FCNC $b\to s + \bar\ell\ell$ decay at very low $q^2 \lesssim 1$ GeV$^2$. The
narrow $(\bar{c}c)$-resonances $J/\psi$ and $\psi'$ are vetoed in the
experimental analysis, while the broader $(\bar{c}c)$-resonances appear at $q^2 >
14$~GeV$^2$.

This singles out a low-$q^2$ and a high-$q^2$ window which are usually chosen in
the intervals $q^2 \in [1, 6]$~GeV$^2$ and $q^2 > 14$~GeV$^2$,
respectively. They correspond to large and low recoil of the $K^*$-meson, as
indicated by the scaling of its energy in the $B$-meson rest frame: $E_{K^*}
\sim m_b/2$ versus $E_{K^*} \sim m_{K^*} + \Lambda_{\rm QCD}$. The heavy quark
limit is combined at large recoil with the large energy limit as an application
of QCD factorisation (QCDF) \cite{Beneke:2001at, Beneke:2004dp} whereas at low
recoil an operator product expansion (OPE) can be performed
\cite{Grinstein:2004vb, Beylich:2011aq} in combination with HQET form factor
relations \cite{Grinstein:2002cz}. Revealing the symmetries of QCD dynamics in
both regions, the expansions reduce the number of hadronic matrix elements. Many
works have been inspired in the past decade in order to exploit these symmetries
to reduce the hadronic uncertainties in the exclusive decays. Particularly, the
angular analysis of the 4-body decay $B \to V (\to P_1 P_2) + \bar\ell\ell$
proved most fruitful. For comparison, $B \to P + \bar\ell\ell$ decays offer only
two additional observables beyond the branching ratio, the lepton
forward-backward asymmetry $A_{\rm FB}^\ell$ and the flat term $F_H^\ell$
\cite{Bobeth:2007dw}.

%
%
\section{$\bf \bar{B}_d^0 \to \bar{K}^{0*} (\to K^- \pi^+) +
  \bar\ell\ell$ \label{sec:Bd-decay}}

\vskip0.3cm

The experimental situation in regard to lepton final states is twofold: On the
one hand, the (Super-) $B$-factories are able to measure both $\ell = e$ and
$\mu$ modes separately. On the other hand, most detectors at hadronic collider
machines prefer $\ell = \mu$, however, LHCb might be able to study $\ell = e$,
too \cite{LHCb-PUB-2009-008}.

%
\subsection{Low-$q^2$ / Large recoil region}

\vskip0.3cm

Using QCDF in the low-$q^2$ region, the seven $B\to V$ QCD form factors ($V,
A_{0,1,2}, T_{1,2,3}$) reduce to two universal form factors
$\xi_{\perp,\parallel}$ \cite{Beneke:2000wa}. Furthermore, the $B \to K^* +
\bar\ell\ell$ amplitudes factorise \cite{Beneke:2001at,
  Beneke:2004dp}\footnote{One might use equivalently soft collinear effective
  theory (SCET) \cite{Ali:2006ew, Lee:2006gs}.}, where the numerically leading
contributions to the transversity amplitudes are \cite{Kruger:2005ep}
\begin{align}
  A_{\perp, \parallel}^{L,R} & \propto \pm c_{\perp}^{L,R} \times \xi_\perp, &
  A_0^{L,R} & \propto c_{\parallel}^{L,R} \times \xi_\parallel,
\end{align}
with short-distance coefficients $c_{\perp,\parallel}^{L,R}$. The transversity
amplitudes have been used to construct the observables $A_T^{(2)}$
\cite{Kruger:2005ep} and $A_T^{(3,4,5)}$ \cite{Egede:2008uy, Egede:2010zc} with
reduced hadronic uncertainties and an improved sensitivity to the chirality
flipped operators ${\cal O}'_{7,9,10}$. Instead of fitting the observables
$J_i^j$ from the angular distribution, the authors of reference
\cite{Egede:2008uy, Egede:2010zc} pursue the possibility to fit directly the
transversity amplitudes which necessitates the identification of their rephasing
properties. Moreover, the authors investigate the experimental reconstruction
uncertainties of their method at LHCb, including among the theoretical
uncertainties also lacking subleading QCDF contributions by simple power
counting.

The original analysis of the CP asymmetries of the rate $A_{\rm CP}$ and the
angular coefficients \cite{Kruger:1999xa}
\begin{align}
  \label{eq:diff:CPasy:2}
  A_i & \equiv \frac{2 \big(J_i^{j} - \bar{J}_i^{j}\big)}
      {d(\Gamma + \bar{\Gamma})/dq^2} ~~\mbox{for}~ i = 3,6,9, &
  A_i^D & \equiv \frac{- 2 \big(J_i^{j} - \bar{J}_i^{j}\big)}
      {d(\Gamma + \bar{\Gamma})/dq^2} ~~\mbox{for}~ i = 4,5,7,8
\end{align}
based on naive factorisation was extended using QCDF to the low-$q^2$ region by
taking into account the NLO QCD corrections for the strong phase $\delta_s$
\cite{Bobeth:2008ij}. Both $A_{3,9}$ vanish in the SM at leading order in QCDF
since they are proportional to the interference of SM and chirality flipped
operators, making them ideal probes of the latter. Furthermore, $A_{7,8}^{D}$
are subject to cancellations at leading order in QCD, such that the NLO QCD
corrections give large corrections. The $q^2$-integrated SM predictions of the
CP asymmetries are summarised in table \ref{tab:loQ2:CPA:SM}. They are, as
expected, tiny and below a percent. A model-independent analysis in the presence
of the BSM contributions to the SM and the chirality flipped operators ${\cal
  O}'_{7,9,10}$ shows substantial room for enhancement. The T-odd asymmetries
$A_{7,8,9}^{D}$ can be of order one. However, the experimental uncertainties at
LHCb are large for CP-asymmetries, even when considering improved normalisations
as discussed for $A_{6s}^{V2s}$ and $A_8^V$ \cite{Egede:2010zc}.

\begin{table}
  \caption{\label{tab:loQ2:CPA:SM} The SM predictions of the $q^2$-integrated
    CP-asymmetries of $\bar{B}_d^0 \to \bar{K}^{0*} (\to K^- \pi^+) +
    \bar\ell\ell$ decays ($q^2 \in [1, 6]$~GeV$^2$) and their
    relative uncertainties from the form factors $\xi_{\perp,\parallel}$ and 
    the renormalisation scale $\mu_b$ \cite{Bobeth:2008ij} (see also 
    \cite{Altmannshofer:2008dz}).
    Also shown are the ranges of the CP asymmetries after applying the
    experimental constraints at 90 \% C.L.~for the generic model-independent
    BSM scenario \cite{Bobeth:2008ij}.
    Note that the SM predictions are given in units of $10^{-3}$.
  }
\begin{center}
\resizebox{\textwidth}{!}{%
\begin{tabular}{c c cccccccc}
\br
&
& $\langle A_{\rm CP} \rangle$ & $\langle A_3 \rangle$ & $\langle
A_4^D \rangle$ & $\langle A_5^D \rangle$
& $\langle A_6 \rangle$ & $\langle A_7^D \rangle$ & $\langle A_8^D
\rangle$ & $\langle A_9 \rangle$
\\
\mr
\multirow{3}{*}{SM}
& $\times 10^{-3}$
& ${4.2}_{ - 2.5}^{ + 1.7}$ & - & ${-1.8}_{ - 0.3}^{ + 0.3}$ &
${7.6}_{ - 1.6}^{ + 1.5}$
& ${-6.4}_{ - 2.7}^{ + 2.2}$ & ${-5.1}_{ - 1.6}^{ + 2.4}$ & ${3.5}_{ -
2.0}^{ + 1.4}$ & -
\\[0.2cm]
& $\xi_{\perp, \parallel} [\%]$
& ${}_{ - 24}^{ + 19}$ & - & ${}_{ - 8}^{ + 11}$ & ${}_{ - 13}^{ + 10}$
& ${}_{ - 39}^{ + 31}$ & ${}_{ - 8}^{ + 11}$ & ${}_{ - 10}^{ + 7.4}$ & -
\\[0.2cm]
& $ \mu_b [\%]$
& ${}_{ - 51}^{ + 33}$ & - & ${}_{ - 6}^{ + 2}$ & ${}_{ - 8}^{ + 7}$
& ${}_{ - 2}^{ + 0}$ & ${}_{ - 26}^{ + 42}$ & ${}_{ - 53}^{ + 37}$ & -
\\
\mr
\multirow{2}{*}{BSM}
& max
& $+0.10$ & $+0.08$ & $+0.04$ & $+0.07$
& $+0.11$ & $+0.76$ & $+0.48$ & $+0.60$
\\
& min
& $-0.12$ & $-0.08$ & $-0.04$ & $-0.07$
& $-0.13$ & $-0.76$ & $-0.48$ & $-0.62$
\\
\br
\end{tabular}
}
\end{center}
\end{table}

The rate-normalised CP-averaged quantities $S_i^{j} \sim (J_i^{j} +
\bar{J}_i^{j})$ constitute another important set of observables, of which the
CP-odd ones ($i = 5,6,8,9$) require $B$-tagged samples. The $S_i^{j}$ have been
analysed in the SM, model-independently and model-dependently in great detail
\cite{Altmannshofer:2008dz, Alok:2010zd}. Especially, the lepton
forward-backward asymmetry $A_{\rm FB}$ (related to $S_6^s$) and $S_5$ could be
measured rather precisely with early LHCb data of $2$~fb$^{-1}$
\cite{Bharucha:2010bb}.

It should be noted that \cite{Egede:2008uy, Egede:2010zc} estimate uncertainties
due to unknown subleading ${\cal O}(\Lambda_{\rm QCD}/E_{K^*})$ contributions to
both form factors and $B \to K^* + \bar\ell\ell$ amplitudes by power counting.
Contrary, \cite{Altmannshofer:2008dz} does not make use of the large recoil
symmetry relations of the form factors in the LO QCD contribution of the said
amplitudes. Instead, all seven QCD form factors determined from light cone sum
rules (LCSR) \cite{Ball:2004rg} are used, thus accounting for the associated
unknown subleading ${\cal O}(\Lambda_{\rm QCD}/E_{K^*})$ contributions. The
direct application of the LCSR results reduces the form factor uncertainties
\cite{Altmannshofer:2008dz} due to more efficient cancellations.

Recently, soft-gluon emission effects due to $(\bar{c}c)$-resonances were found
to extend down to the low-$q^2$ region \cite{Khodjamirian:2010vf} affecting the
$q^2$-differential rate at most about $+15$ \% at $q^2 = 6$ GeV$^2$. Due to this
relatively large shift such effects should be investigated further and should be
considered in future studies of the exclusive rare decays.

%
\subsection{High-$q^2$ / Low recoil region}

\vskip0.3cm

\begin{figure}
\begin{minipage}[t]{0.49\textwidth}
  \includegraphics[width=3.8cm, angle=-90]{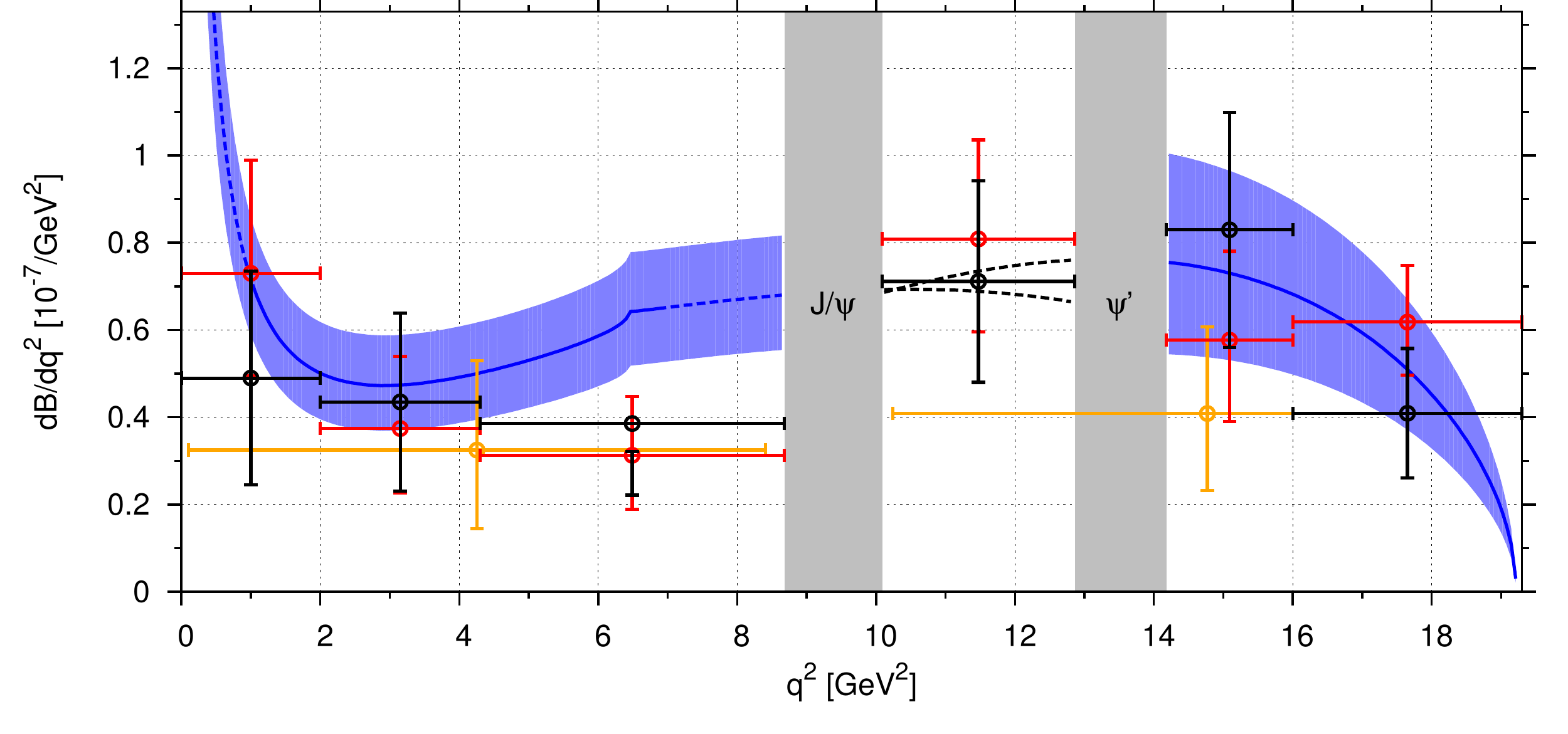} \\[0.3cm]
  \includegraphics[width=3.8cm, angle=-90]{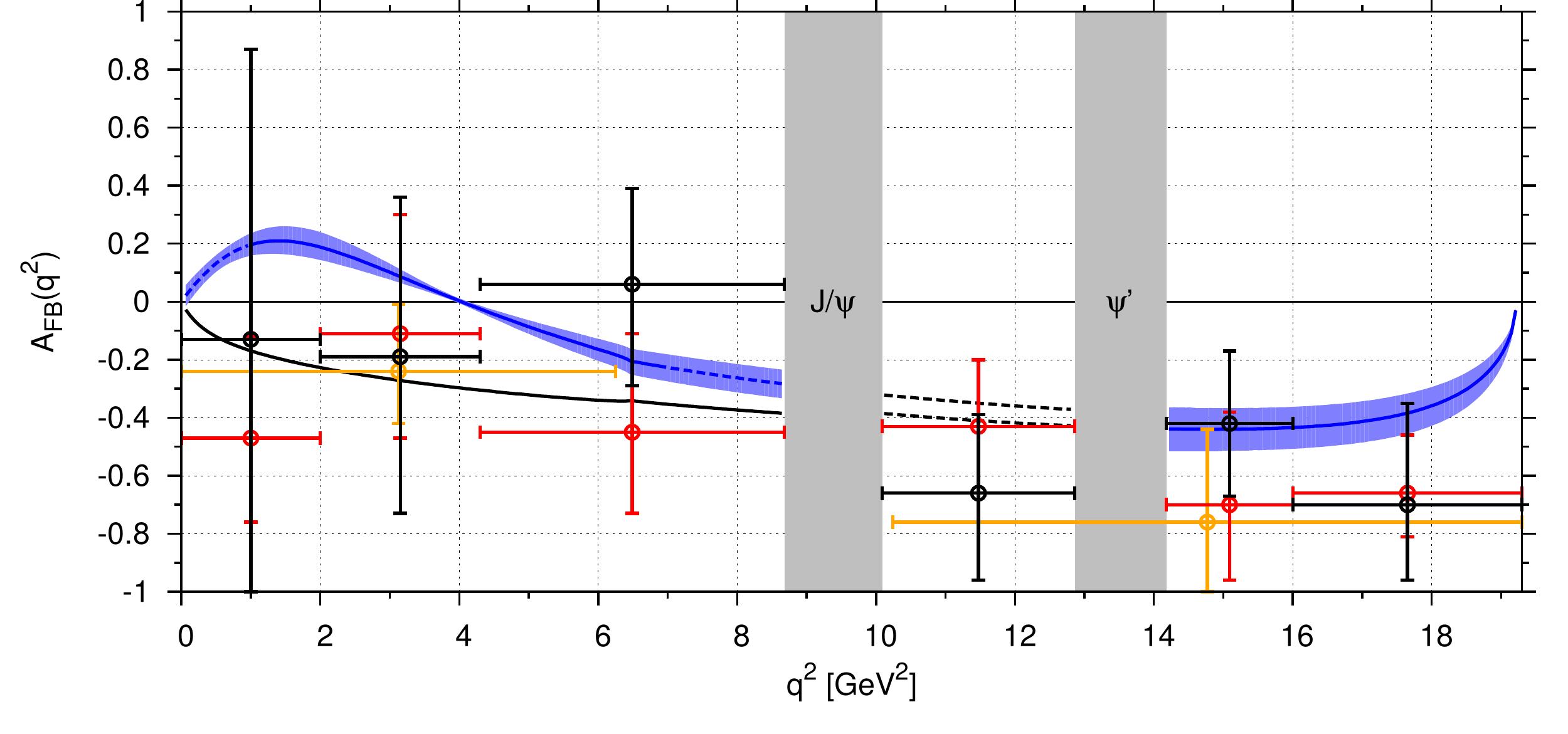}
\end{minipage}
\hspace{0.02\textwidth}
\begin{minipage}[t]{0.49\textwidth}
  \includegraphics[width=3.8cm, angle=-90]{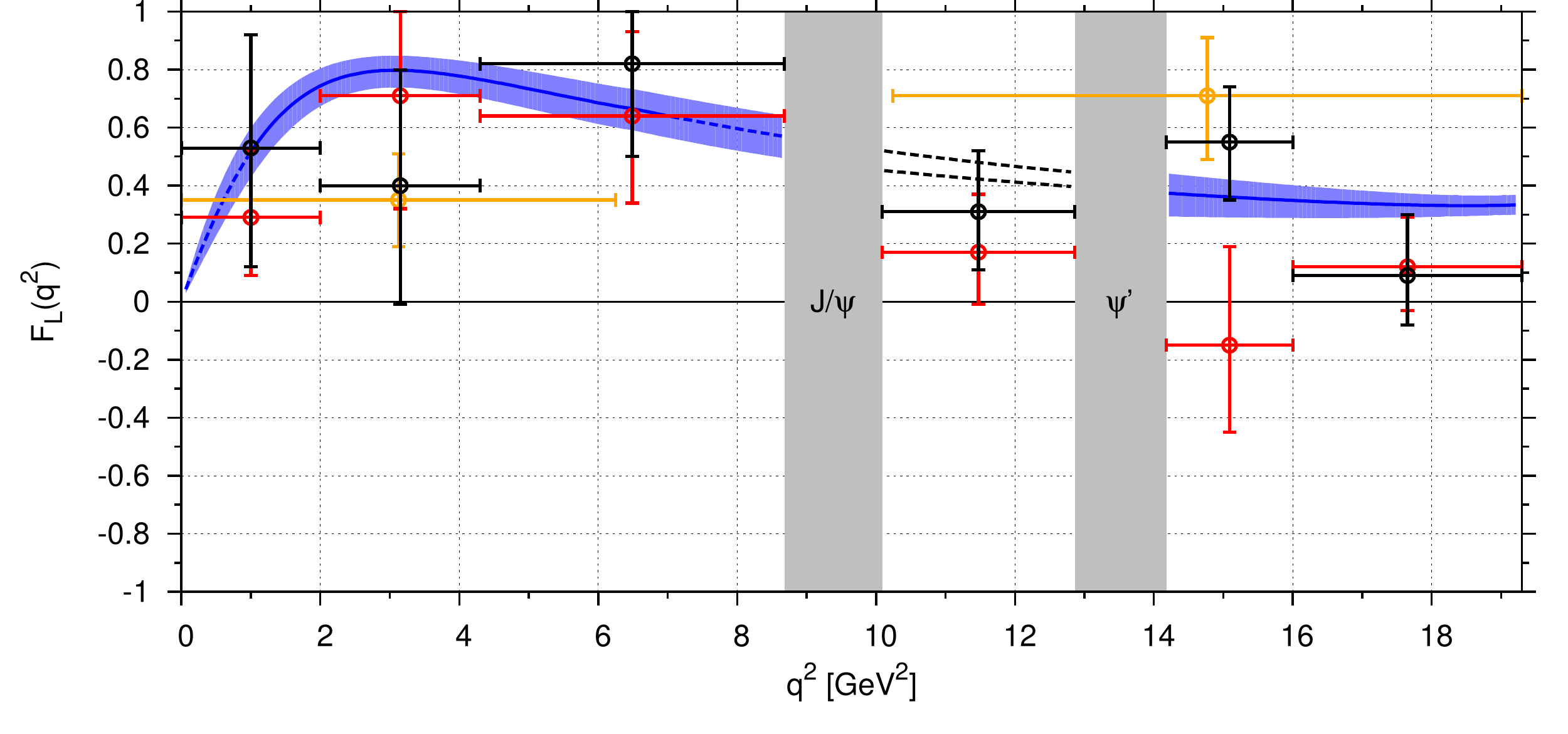} \\
  \caption{\label{fig:q2-distr} Comparison of BaBar [yellow], Belle [red] and
    CDF [black] data points with SM predictions (blue) for $\bar{B}_d^0 \to
    \bar{K}^{*0} + \bar\ell\ell$ distributions \cite{Bobeth:2010wg}: $d{\cal
      B}/dq^2$, $A_{\rm FB}$ and $F_L$. The vertical grey bands are vetoed in
    experiment.}
\end{minipage}
\end{figure}

Assuming quark-hadron duality, the low recoil region can be studied using an OPE
in combination with HQET to treat the long-distance contributions from
current-current and QCD-penguin operators \cite{Grinstein:2004vb,
  Beylich:2011aq}. The authors of \cite{Beylich:2011aq} perform the OPE without
subsequent matching to HQET contrary to \cite{Grinstein:2004vb}, among further
technical details. Duality violating corrections to the OPE are estimated in
\cite{Beylich:2011aq} and found to be small. In \cite{Grinstein:2004vb} it is
shown, that at the leading order in the $\Lambda_{\rm QCD}/Q$, $Q =
\{\sqrt{q^2},m_b\}$ expansion\footnote{Note, that the subleading order is
  additionally suppressed by $\alpha_s$.} all hadronic matrix elements can be
expressed through the 3 QCD form factors $V, A_{1,2}$. As a consequence, a
universal short-distance dependence emerges for all transversity amplitudes ($a
= 0, \perp, \parallel$) ~\cite{Bobeth:2010wg}
\begin{align}
  \label{eq:hiQ2:fac}
  A_a^{L,R} &\propto c_{L,R} \times f_a, &
  c_{L,R} & = ({\cal C}_9^{\rm eff} \mp {\cal C}_{10})
           + \kappa \frac{2 m_b}{q^2} {\cal C}_7^{\rm eff},
\end{align}
with the (effective) short-distance couplings ${\cal C}_{7,9,10}^{\rm eff}$ and
the form factor terms $f_a$ ($a = 0, \perp, \parallel)$ which are linear in $V,
A_{1,2}$. The factor $\kappa \sim 1$ is known including NLO in QCD from matching
QCD onto HQET \cite{Grinstein:2004vb}. The term $\propto {\cal C}_7^{\rm eff}$
is subject to uncertainties of order $\Lambda_{\rm QCD}/m_b$ due to the HQET
form factor relations, which are additionally suppressed in the SM by $|{\cal
  C}_7/{\cal C}_9| \sim 0.1$. As a consequence of (\ref{eq:hiQ2:fac}) $J_7 = J_8
= J_9 = 0$ and the remaining angular observables
\begin{align}
  J_2^c & \sim U_1 = 2 \rho_1 f_0^2, &
  (2 J_2^s + J_3) & \sim U_2 = 2 \rho_1 f_\perp^2, &
  (2 J_2^s - J_3) & \sim U_3 = 2 \rho_1 f_\parallel^2,
  \label{eq:hiQ2:I}
\\ 
  J_4 & \sim U_4 = 2 \rho_1 f_0 f_\parallel, &
  J_5 & \sim U_5 = 4 \rho_2 f_0 f_\perp, &
  J_6^s & \sim U_6 = 4 \rho_2 f_\parallel f_\perp, &
  \nonumber
\end{align}
depend only on the two short-distance (dominated) combinations $\rho_1 =
(|c_R|^2 + |c_L|^2)/2$ and $\rho_2 = (|c_R|^2 - |c_L|^2)/4$
\cite{Bobeth:2010wg}. From (\ref{eq:hiQ2:I}) one obtains the three long-distance
free ratios
\begin{align}
  H_T^{(1)} & = \frac{U_4}{\sqrt{U_1 \cdot U_3}} = 1, & 
  H_T^{(2)} & = \frac{U_5}{\sqrt{U_1 \cdot U_2}} = 2\, \frac{\rho_2}{\rho_1}, &
  H_T^{(3)} & = \frac{U_6}{\sqrt{U_2 \cdot U_3}} = 2\, \frac{\rho_2}{\rho_1}. 
\\
\intertext{After integration over $q^2 \in [14.0 , 19.2]$~GeV$^2$ one obtains 
  in the SM}
  \langle H_T^{(1)} \rangle & = +0.997 \pm 0.003, &
  \langle H_T^{(2)} \rangle & = -0.972 \pm 0.010, &
  \langle H_T^{(3)} \rangle & = -0.958 \pm 0.010.
\end{align} 
Details concerning the definition of $\langle \ldots \rangle$ and the
theoretical uncertainties are given in \cite{Bobeth:2010wg} as well as the
results for ${\cal B}, A_{\rm FB}, F_L$ and $A_T^{(2,3,4)}$. We stress that
$\langle H_T^{(1)} \rangle \approx 1$ holds model-independently. Hence,
deviations from this prediction test the validity of the OPE
framework. Furthermore, the following short-distance free ratios
\begin{align}
  \frac{f_0}{f_\parallel} & = \sqrt{\frac{U_1}{U_3}} = \frac{U_1}{U_4} 
    = \frac{U_4}{U_3} = \frac{U_5}{U_6}, &
  \frac{f_0}{f_\perp} & = \sqrt{\frac{U_1}{U_2}},  &
  \frac{f_\perp}{f_\parallel} & = \sqrt{\frac{U_2}{U_3}} 
    = \frac{\sqrt{U_1 U_2}}{U_4}
\end{align}
 allow to test
the $B \to V$ form factors at high-$q^2$, such as obtained from lattice QCD,
using experimental data \cite{Bobeth:2010wg}.

\begin{figure}
\begin{minipage}[b]{0.70\textwidth}
  \includegraphics[width=5.5cm, angle=0]{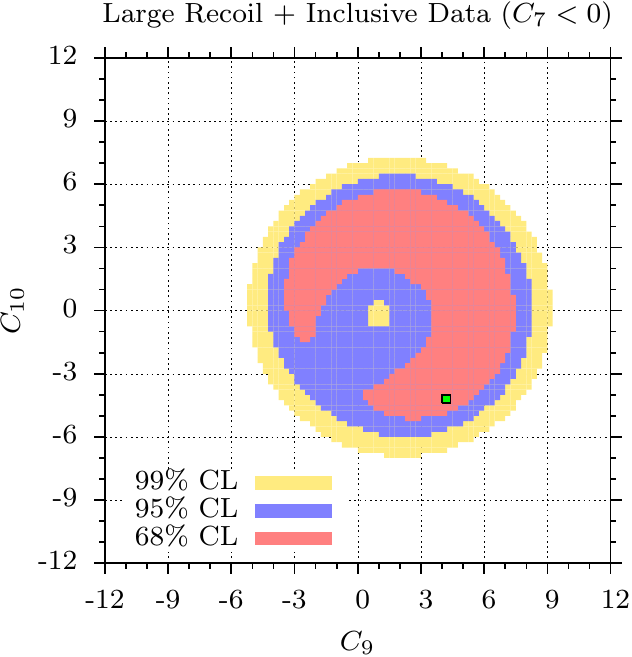}
  \includegraphics[width=5.5cm, angle=0]{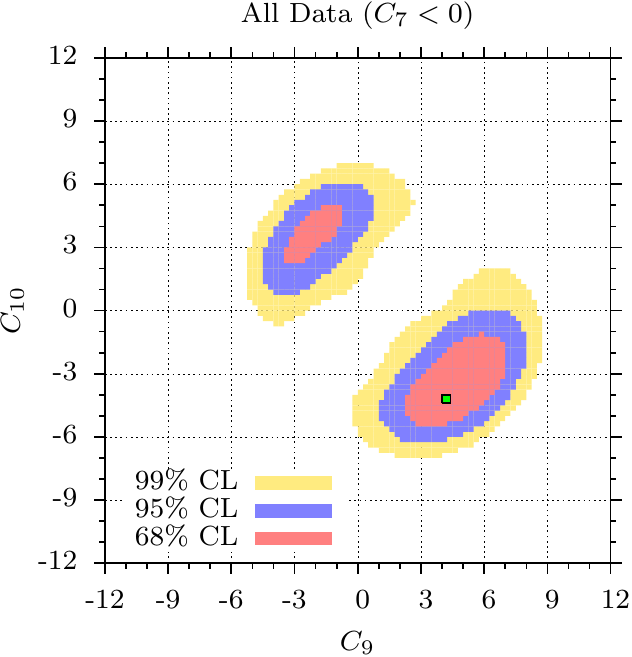}
\end{minipage}
\hspace{0.01\textwidth}
\begin{minipage}[b]{0.29\textwidth}
  \caption{\label{fig:c9c10-real} Constraints on real-valued ${\cal C}_{9,10}$
    at 68\%, 95\% and 99\% CL from low-$q^2$ bins of $\bar{B}_d^0 \to
    \bar{K}^{*0} \bar\ell\ell$ and $B\to X_s \bar\ell\ell$ [left] and when
    adding high-$q^2$ bins [right] \cite{Bobeth:2010wg, Bobeth:2011gi}. Here
    ${\cal C}_{7} = + {\cal C}_{7}^{\rm SM}$ is fixed. The square marks the SM
    value of (${\cal C}_{9}, {\cal C}_{10}$).}
\end{minipage}
\end{figure}

\begin{figure}
\begin{minipage}[b]{0.70\textwidth}
  \includegraphics[width=5.5cm, angle=0]{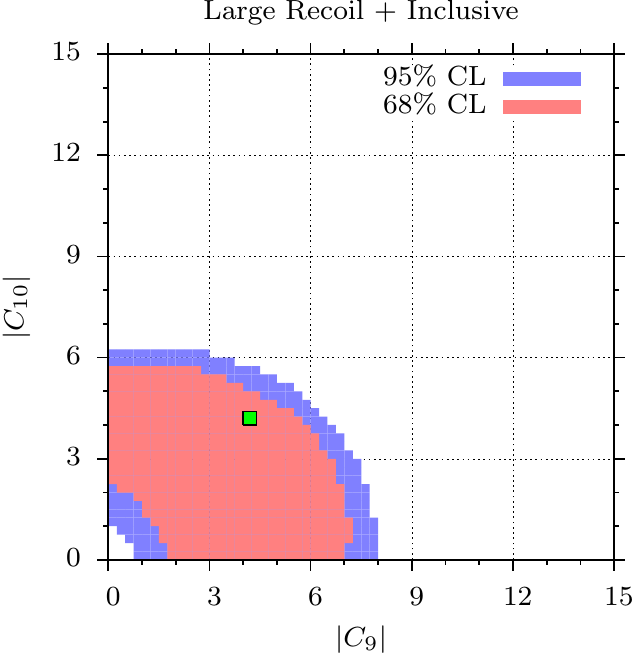}
  \includegraphics[width=5.5cm, angle=0]{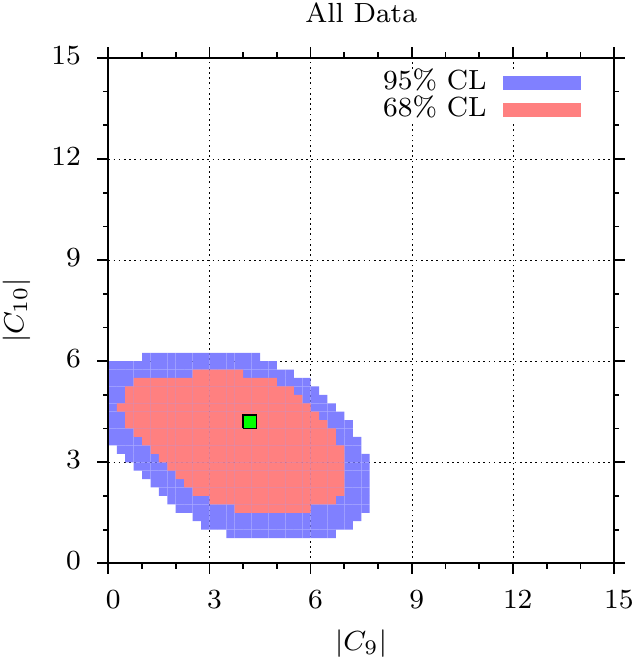}
\end{minipage}
\hspace{0.01\textwidth}
\begin{minipage}[b]{0.29\textwidth}
  \caption{\label{fig:c9c10-cplx} Constraints on $|{\cal C}_{9,10}|$ at 68\% and
    95\% CL for complex-valued ${\cal C}_{9,10}$ from low-$q^2$ bins of
    $\bar{B}_d^0 \to \bar{K}^{*0} \bar\ell\ell$ and $B\to X_s \bar\ell\ell$
    [left] and when adding high-$q^2$ bins [right] (see \cite{Bobeth:2011gi} for
    details). The square marks the SM value of (${\cal C}_{9}, {\cal C}_{10}$).
  }
\end{minipage}
\end{figure}

A comparison of the SM predictions and the available data from BaBar, Belle and
CDF for low- and high-$q^2$ regions is shown in figure \ref{fig:q2-distr} where
the main uncertainties (shaded bands) arise from the form factors and the
missing subleading corrections of order $\Lambda_{\rm QCD}/m_b$ as estimated by
power counting \cite{Bobeth:2010wg}. The Belle and CDF measurements of the three
low- and high-$q^2$ bins $q^2 \in [1, 6], [14.18, 16.0], [16.0, 19.21]$~GeV$^2$
of ${\cal B}$ and $A_{\rm FB}$ (and $q^2 \in [1, 6]$~GeV$^2$ of $F_L$) yield the
first constraints on the short-distance couplings ${\cal C}_{9,10}$ of the SM
operators ${\cal O}_{9,10}$ beyond naive factorisation. The constraints for
real-valued ${\cal C}_{9,10}$ are shown in figure \ref{fig:c9c10-real} fixing
${\cal C}_{7} = + {\cal C}_{7}^{\rm SM}$ (results for ${\cal C}_{7} = - {\cal
  C}_{7}^{\rm SM}$ are available in \cite{Bobeth:2010wg}) when employing
low-$q^2$ data only and when adding high-$q^2$ data\footnote{Using an
  extrapolation of the $B\to V$ form factors from low-$q^2$ QCD LCSR
  \cite{Ball:2004rg}.}, respectively. Figure \ref{fig:c9c10-cplx} shows $|{\cal
  C}_{10}|$ vs. $|{\cal C}_{9}|$ for complex-valued ${\cal C}_{9,10}$ and
results of ($\phi_9$ vs. $|{\cal C}_{9}|$), ($\phi_{10}$ vs. $|{\cal C}_{10}|$)
and ($\phi_{10}$ vs. $\phi_9$) can be found in \cite{Bobeth:2011gi}. Both
figures demonstrate the strong constraining power of the high-$q^2$ region on
${\cal C}_{9,10}$, and consistency with low-$q^2$ data.

CP asymmetries with improved short-distance sensitivity can be designed in
analogy to $H_T^{(1,2,3)}$, since $\bar{J}_i^j \sim \bar{\rho}_a$ with
$\bar{\rho}_a = \rho_a [\delta_W \to - \delta_W]$. It is convenient to consider
the three CP-asymmetries
\begin{align}
 a_{\rm CP}^{(1)} & = \frac{\rho_1 - \bar{\rho}_1}{\rho_1 + \bar{\rho}_1}, &
 a_{\rm CP}^{(2)} & = \frac{\frac{\rho_2}{\rho_1} - \frac{\bar{\rho}_2}{\bar{\rho}_1}}
                          {\frac{\rho_2}{\rho_1} + \frac{\bar{\rho}_2}{\bar{\rho}_1}}, &
 a_{\rm CP}^{(3)} & = 2\, \frac{\rho_2 - \bar{\rho}_2}{\rho_1 + \bar{\rho}_1}, &
\end{align}
with various possibilities to measure them \cite{Bobeth:2011gi}. Moreover,
$a_{\rm CP}^{(3)}$ can be extracted from an untagged $B$-meson sample. There are
no T-odd CP-asymmetries at low-recoil due to (\ref{eq:hiQ2:I}): $J_{7,8,9} = 0$.
In the SM $|a_{\rm CP}^{(1,2,3)}|_{\rm SM} \lesssim 10^{-4}$ whereas a
model-independent analysis for complex-valued ${\cal C}_{9,10}$ allows values up
to $|\langle a_{\rm CP}^{(1,3)} \rangle| \lesssim 0.2$ and $|\langle a_{\rm
  CP}^{(2)} \rangle|$ is presently unconstrained (see \cite{Bobeth:2011gi} for
details).

%
%
\section{CP asymmetries without tagging \label{sec:Bs-CPAs}}

\vskip0.3cm

The decays $\bar{B}_s,B_s \to \phi (\to K^- K^+) + \bar\ell\ell$ require to
account for the mixing of the initial $\bar{B}^0_s$- and $B^0_s$-mesons before
decaying into the common final state. Consequently, their observables depend
additionally on the $\bar{B}_s - B_s$ mixing phase $\Phi_s$ and the lifetime
difference $\Delta\Gamma_s$, which are currently being analysed at the Tevatron,
for instance in $\bar{B}_s,B_s \to \phi (\to K^+ K^-) + J/\psi (\to
\bar\ell\ell)$. In this respect, the CP-odd property of the observables $J_i^j$
with $i = 5,6,8,9$ allows to measure the associated CP-asymmetries from an
untagged and time-integrated data set.

The CP-asymmetries $A_i^{mix}$ for $i=6,9$ and $A_i^{D mix}$ for $i = 5,8$ have
been worked out at low-$q^2$ \cite{Bobeth:2008ij}. They match the corresponding
$A_i$ and $A_i^{D}$ of the flavour-specific unmixed decays for $\Delta\Gamma_s
\to 0$. There the CP-averaged rate has been used as normalisation. An improved
cancellation of the hadronic uncertainties might be achieved along the lines of
the unmixed case by choosing the normalisations proposed for $A_{6s}^{V2s}$ and
$A_8^V$ \cite{Egede:2010zc}.
 
At high-$q^2$ the CP-asymmetries associated with $J_i^j$, $i=8,9$ vanish and
those with $i = 5, 6$ can be related to the same short-distance dominated
$a_{\rm CP}^{mix}$ using the normalisations as indicated in reference
\cite{Bobeth:2011gi}. Notably, $a_{\rm CP}^{mix}$ is independent of the sign of
$\Delta\Gamma_s$ and the dependence on $\Phi_s$ is rather weak for small
$\Delta\Gamma_s/\Gamma_s$. Experimentally, $\Delta\Gamma_s/\Gamma_s \sim {\cal
  O} (0.1)$.

%
%
\section{Conclusion \label{sec:concl}}

\vskip0.3cm

The angular analysis of the exclusive $\bar{B}_d^0 \to \bar{K}^{0*} (\to K^-
\pi^+) + \bar\ell\ell$ and $\bar{B}_s,B_s \to \phi (\to K^- K^+) + \bar\ell\ell$
decays with 4-body final states offers a rich phenomenology to test the
short-distance flavour couplings in a large number of observables. The LHCb
experiment is expected to improve the experimental situation tremendously in the
near future, but also the upcoming final analysis of the complete BaBar, Belle
and CDF data sets will contribute to the exploration of the borders of the SM.

The low- and high-$q^2$ regions are accessible to power expansions which reveal
symmetries of QCD and indicate suitable combinations of observables to reduce
the hadronic uncertainties. This allows for precise theory predictions for the
exclusive decays.

The low-$q^2$ region is theoretically (QCDF, SCET) well understood and effects
of $(\bar{c}c)$-resonances can be estimated. It offers many interesting
observables which wait for confrontation with data.

The high-$q^2$ region is based on an OPE relying on quark-hadron duality.
Violations of the latter could be signaled experimentally by deviations from
$H_T^{(1)} = 1$. Currently, the $B \to V$ form factors used at high-$q^2$ are
extrapolated from the low-$q^2$ region using a dipole formula, but lattice
calculations for the high-$q^2$ regions are in progress \cite{Liu:2009dj}. There
are two ``long-distance free'' ratios $H_T^{(2,3)}$ which provide tests of the
SM and, moreover, several ``short-distance free'' ratios which offer direct
comparison of the lattice results for ratios of form factors with data.

A new flavour tool ``EOS'' aimed at the evaluation of the observables covered in
this article is developed at TU Dortmund \cite{EOS:2011} whose first stable
release is expected during 2011.

%
%
\ack 

\vskip0.3cm

C.~B. thanks the organisers of the {\em DISCRETE 2010 - Symposium on Prospects in
  the Physics of Discrete Symmetries} for the opportunity to present a talk and
the kind hospitality in Rome.  The work reported here has been supported in part
by the Bundesministerium f\"ur Bildung und Forschung (BMBF).

%
%

\end{document}